\documentclass[twocolumn,aps,prl,groupedaddress,amssymb,showpacs]{revtex4}
\usepackage[dvips]{graphicx,color}
\usepackage{natbib}
\begin{document}
 \definecolor{blue}{rgb}{0.24,0.24,1.00}
  \definecolor{red1}{rgb}{1.000000,0.000000,0.000000}
\title{A comparison of the magnetic properties of Proton- and Iron-implanted graphite}
\author{J. Barzola-Quiquia}
\author{R. H\"ohne}
\author{M. Rothermel}
\author{A. Setzer}
\author{P. Esquinazi}\email{esquin@physik.uni-leipzig.de}
\affiliation{Institut f\"{u}r Experimentelle Physik II,
Universit\"{a}t Leipzig, Linn\'{e}stra{\ss}e 5, D-04103 Leipzig,
Germany}
\author{V. Heera}
 \affiliation{Institute of Ion Beam Physics and Materials Research, Forschungszentrum Dresden-Rossendorf,
 P.O.Box 510119, D-01314 Dresden, Germany} %%
\begin{abstract}
In this work we have investigated the changes of the magnetic
properties of highly oriented pyrolytic graphite samples after
irradiation either with $\sim 3 \times 10^{14}$ protons or $ 3.5
\times 10^{13} \ldots 3.5 \times 10^{14}$ iron ions  with energies in
the MeV range. Our results show that iron and proton irradiations can
produce similar paramagnetic contributions depending on the
implantation temperature. However, only protons induce a
ferromagnetic effect.
\end{abstract}
 \maketitle

The possible existence of ferromagnetic order in metal-free carbon
attracted some interest of basic and material research scientists
since the end of the 60's\cite{magata,tyu74,ovchi78,maka04}. However,
only in the last years and partially due to the technical improvement
in the magnetic detection as well as in the analytical methods to
determine the impurity contribution to the magnetic properties, this
subject shows nowadays a renaissance. Theoretical studies showed that
 a mixture of sp$^2$-sp$^3$-bonded carbon atoms with unpaired
$\pi$-electrons \cite{ovchi91} or different numbers of mono- and
dihydrogenated carbon atoms \cite{kusakabe03}, both within a
graphite-like structure, may trigger magnetic order. New
theoretical approaches suggest, however, that lattice defects
and/or the influence of atomic hydrogen coupled to a carbon atom
of the graphene lattice might induce a spontaneous magnetization
\cite{gon01,lehtinen04,duplock04,yaz07}. Depending on the density
and position of the defects, different Curie temperatures are
expected. The influence of an hydrogen-like particle on the
magnetic response of the graphite lattice has been indirectly
observed by muon spin resonance ($\mu$SR) experiments on highly
oriented pyrolytic graphite (HOPG) \cite{cha02}, which reveal the
formation of a magnetic moment in the surrounding of the muon. It
is important to note that in the theoretical approaches the
graphite/graphene lattice plays a main role, i.e. strictly
speaking amorphous carbon or a highly disordered graphite lattice
is not expected to reveal spontaneous magnetization. Whether the
ferromagnetic carbon films reported in
Refs.~\onlinecite{murata91,murata92,mizogami91} were highly
disordered or amorphous remains still unclear since apparently
nobody could reproduce those results yet.

Whereas earlier obtained experimental evidence for room
temperature magnetic order in graphite
\cite{yakovjltp00,pabloprb02,pabloprl03,esq05} has been recently
confirmed \cite{ohldagl,barzola1,barzola2}, part of the scientific
community still doubts concerning the intrinsic character of the
observed effects added to a special role of iron in a carbon
matrix \cite{coey02,mertins04}. The aim of the present study was
to compare the magnetic properties of HOPG samples bombarded by a
similar number of protons- and iron-ions at similar energies. Our
experimental work shows that iron bombardment under the used
irradiation conditions and up to a concentration of 265 ppm (in
the irradiated region) does not trigger magnetic order in
graphite, in clear contrast to proton irradiation.

In these studies two main improvements were realized in comparison
with earlier reports on the measurements of the magnetic properties
of proton irradiation of HOPG samples \cite{pabloprl03,esq05}.
Firstly, in order to reduce the effective thermal annealing of the
defects produced during irradiation an special chamber was developed
that enables us to cool down the sample at a nominal temperature of
110~K during irradiation. Second, a special sample holder was
designed that allows us to measure the sample in the SQUID and to fix
it inside the irradiation chamber without any changes
\cite{barzola1}. This last improvement provides a reproducibility in
magnetic moment of $\sim 10^{-7}$~emu. In the case of iron
implantation the samples must be every time fixed on the holder for
SQUID measurements after irradiation. Therefore, small deviations
exist in the sample orientation respect to the applied magnetic field
between measurements performed before and after irradiation, which
influences mainly the measured diamagnetic contribution. This
contribution should be accordingly subtracted in order to obtain any
para- and/or ferromagnetic contributions.

For the present investigations HOPG samples of grade ZYA
($0.4^\circ$ mosaicity) were used. To trigger ferromagnetism by
proton irradiation of HOPG broad- as well as micro-beam technique
can be used \cite{barzola2}. The sample HG1 was irradiated at
110~K with a 2.25 MeV proton microbeam perpendicular to the
graphite planes. The irradiated sample is called HG1$_-$H (the
same sample before irradiation is called HG1$_-$V). The
irradiation was performed on $160 \times 160$ spots of $\simeq
2.6~\mu$m radius each and separated by a distance of $10~\mu$m on
a total area of 2.5~mm$^2$. The dose (fluence) and total charge
were $5 \times 10^{16}$~cm$^{-2}$ (0.08~nC/$\mu$m$^2$) and
44.8~$\mu$C, i.e. a total of $\simeq 2.8 \times 10^{14}$~protons
up to a depth of $\sim 46~\mu$m from the surface. The proton
current used was 1~nA. Using particle induced x-ray emission we
determined a magnetic impurity content of $0.4~\mu$g/g of Fe ($<
0.1~$ppm) homogeneously distributed in the sample.

Sample HG2$_-$V was implanted with Fe-ions onto an area of
6.9~mm$^2$ up to a depth of $\simeq 2.1~\mu$m by multi-energy
implantation in the range between $1\ldots 4.0~$MeV. The sample
was implanted with a fluence of $5 \times 10^{14}~$Fe/cm$^{2}$
(sample HG2$_-$0.5Fe). After magnetic characterization, the same
sample was irradiated at $450^\circ$C with a fluence of $5 \times
10^{15}~$Fe/cm$^2$ on the same surface after renewing it by
peeling out the earlier irradiated surface; this sample is called
HG2$_-$5.0Fe. The achieved Fe concentration in the implanted depth
is 26.5~ppm for sample HG2$_-$0.5Fe and 265~ppm for sample
HG2$_-$5.0Fe. The measurements of the magnetic moment were carried
out with a SQUID magnetometer from Quantum Design with a
reciprocating sample option (RSO) and a resolution of $\sim 2
\times 10^{-8}$emu. Both samples HG1 and HG2 were characterized
magnetically in the virgin as well as in the irradiated states.
The field was applied always parallel to the graphene layers to
reduce the diamagnetic contribution to the magnetic moment.

\begin{figure}[]
%\vspace{0.5cm}
\begin{center}
\includegraphics[width=85mm]{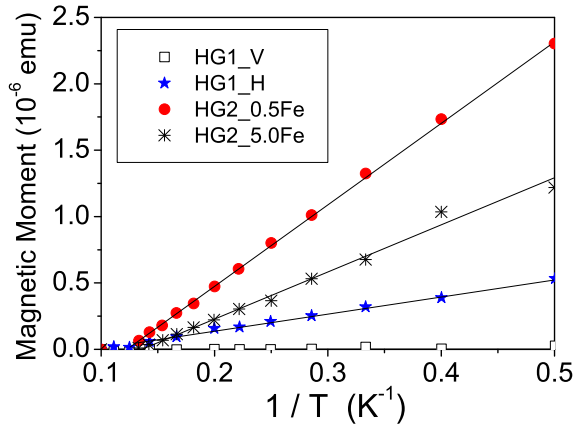}
\caption[]{Magnetic moment in units of $10^{-6}$~emu (1 emu/g = 1
Am$^2$/kg) measured at $H = 2~$kOe (1~Oe = $10^3/4\pi$~A/m) after
ZFC from 300~K to 2~K as a function of the reciprocal temperature
for sample HG1 in the virgin state $(\square)$ and after proton
irradiation $({\color{blue} \bigstar})$ as well as for sample HG2
after the first $({\color{red1} \bullet})$ and the second iron
implantation $(\ast)$. The curves are obtained after subtracting
the magnetic moment at $T = 10$~K and 2~kOe. The straight lines
are linear fits to the data.} \label{fig1}
\end{center}
\end{figure}

In the investigated temperature range we observe three
contributions to the magnetization of the same graphitic phase: a
diamagnetic one due to orbital effects, a paramagnetic one caused
by the lattice disorder before and (largely enhanced) after
irradiation, and a weak ferromagnetic one with a Curie temperature
above 300~K. The paramagnetic contribution is visible in an
up-turn of the $T-$dependence of the magnetic moment at $T
\lesssim 20~$K. Figure~1 shows the magnetic moment $m$ vs. inverse
$T$ after subtraction  of the value $m(10$K) at the applied field
of 2~kOe. Due to the practically $T-$independence of the
ferromagnetic and diamagnetic contributions between 2~K and 10~K,
this subtraction shows basically the paramagnetic contribution,
which follows the $1/T$ Curie-law dependence, see Fig.~1. The
slope of the lines in Fig.~1 is proportional to the density of
paramagnetic centers. The paramagnetic contribution of the sample
in the virgin state (HG1$_-$V) is very small whereas the disorder
created by the particle bombardment enhances this contribution.

We note that the implantation with iron performed at room
temperature shows the highest paramagnetism among all samples. The
paramagnetic contribution induced by iron implantation  at the
same temperature increases with implantation dose. It decreases,
however, by increasing the implantation temperature or by a
post-implantation heat treatment. The sample HG2$_-$5.0Fe was
implanted with the tenfold dose of that of sample HG2$_-$0.5Fe but
at $T = 450~^\circ$C. Because of the strong recombination of the
carbon vacancies created by the Fe bombardment the paramagnetic
contribution is smaller than that in sample HG2$_-$0.5Fe and can
be further decreased by a factor of two after a heat treatment at
$700~^\circ$C in high vacuum for two hours.

\begin{figure}[]
%\vspace{0.5cm}
\begin{center}
\includegraphics[width=85mm]{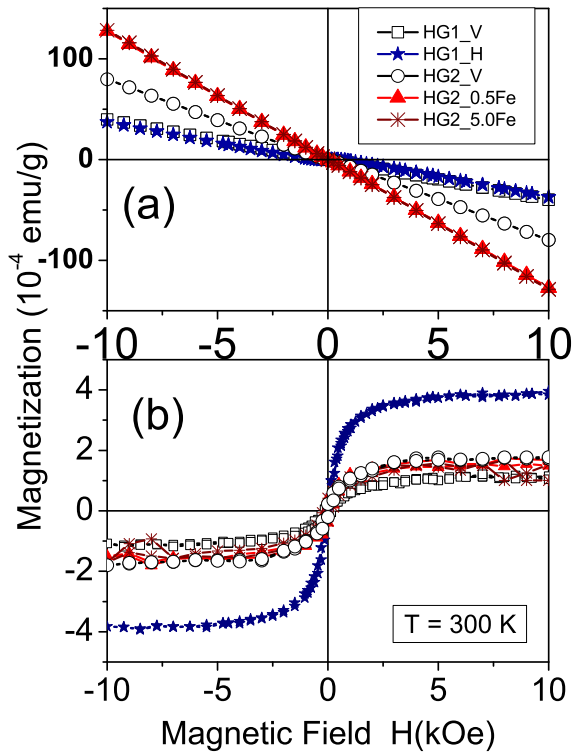}
\caption[]{Magnetization in units of $10^{-4}~$emu/g (related to
the whole sample mass) measured at $T = 300~$K as a function of
the magnetic field for the samples before (open symbols) and after
irradiation with protons $({\color{blue} \bigstar})$ or Fe-ions
$({\color{red1} \blacktriangle},$ sample HG2$_-$0.5Fe, $\ast$,
sample HG2$_-$5.0Fe) without any subtraction (a) and after
subtraction of the linear diamagnetic part (b). The remanent
magnetization and coercive fields for sample HG1$_-$V, for
example, are $\simeq 2.4 \times 10^{-5}~$emu/g and $\simeq
0.16~$kOe.} \label{fig2}
\end{center}
\end{figure}
For a quantitative comparison we estimate the number of Bohr
magnetons $\mu_B$ in the samples using the Curie law and the
assumption that each magnetic center contributes with one $\mu_B$.
From the slopes of the curves in Fig.~1 one gets that the proton
irradiation of sample HG1 has induced $3.2 \times 10^{15}~\mu_B$, the
samples HG2$_-$0.5Fe Fe-implanted at room temperature and
HG2$_-$5.0Fe Fe-implanted at $450~^\circ$C contain $\simeq 14.6
\times 10^{15}~\mu_B$ and $\simeq 8.1 \times 10^{15}~\mu_B$,
respectively. We note that in the sample HG2$_-$5.0Fe a particular
part of the paramagnetic contribution is caused by the paramagnetic
Fe-ions themselves. Let us compare these values in relation to the
number of implanted particles. From the total charge of $44.8~\mu$C
and the charge of an electron one gets $2.8 \times 10^{14}$ implanted
protons in sample HG1$_-$H. Into the two Fe-implanted samples we
implanted $3.45 \times 10^{13}$ and $3.45 \times 10^{14}$ Fe-ions.
From these values follows that the numbers of Bohr magnetons per
implanted particle is  $11 \pm 2, 420 \pm 100$ and $23 \pm 5$ for
samples HG1$_-$H, HG2$_-$0.5 Fe and HG2$_-$5.0 Fe, respectively. The
relatively large (maximal) error of these values for the Fe-implanted
samples  comes from the possible deviations between the nominal and
the realized iron implantation dose and the uncertainties concerning
the paramagnetic contribution of sample HG2 in its virgin state.
Without doubt the maximum disorder is created by the Fe-implantation
at room temperature. This in principle correlates with the results of
SRIM simulations \cite{ziegler}. The integral of the damage
distribution caused by the implantation  under the here used
conditions gives 15 and 4000 carbon displacements (vacancies) per
implanted proton and iron ion. For example, for the sample
HG2$_-$0.5Fe we get a vacancy concentration of $1 \times
10^{22}~$vacancies/cm$^3$ in the first $\sim 2~$nm from the surface.
Compared with the atomic density of carbon in HOPG this value
corresponds to 0.088 displacements per atom.
\begin{figure}[]
%\vspace{0.5cm}
\begin{center}
\includegraphics[width=85mm]{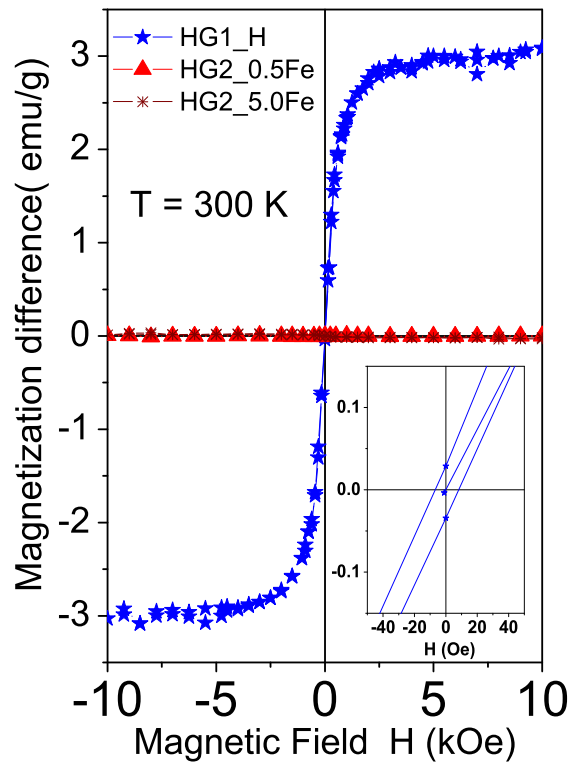}
\caption[]{Magnetization in units of emu/g for the irradiated
samples after subtraction of the contribution of the corresponding
samples in the virgin state measured at $T = 300$~K as a function
of the magnetic field. The magnetization  is related to the
estimated ferromagnetic mass of the irradiated sample region only.
The inset shows the region around zero field of the proton
irradiated sample HG1$_-$H.} \label{fig3}
\end{center}
\end{figure}
However, most of these vacancies are not stable and due to
vacancy-interstitial recombination the damage is reduced by almost
one order of magnitude, i.e. instead of 4000$~\mu_B$ (assuming
$1~\mu_B$ per vacancy) we obtain experimentally $\simeq 420~\mu_B$
per implanted Fe-ion. Nevertheless the lattice disorder remains high
in the Fe-implanted sample HG2$_-$0.5Fe.

For the discussion of the lattice damage we have to take into account
the different defect distribution for iron and proton irradiations.
Whereas this distribution is rather homogeneous in our Fe-implanted
samples (due to the multi-energy implantation method), the 2.25 MeV
proton irradiation shows a sharp defect density maximum at $\sim
46~\mu$m. However, the lattice disorder in the proton-irradiated
sample is relatively small in the ferromagnetically active layer,
which is localized in the first micrometers \cite{barzola2}. In this
region the proton irradiation produced a vacancy density of $\simeq 5
\times 10^{20}~$cm$^{-3}$. We note that this estimated density from
SRIM simulations is 20 times smaller than the one for the
Fe-irradiation of sample HG2$_-$0.5Fe. The ratio 20 between these
densities of vacancies is comparable with the experimental ratio of
40 for the numbers of Bohr magnetons per implanted particle.

The magnetization measured at $T = 300~$K by cycling the magnetic
field between $\pm 10~$kOe is given in Fig.~2(a). For all samples the
magnetization curves are almost linear, dominated by the
diamagnetism. The differences of the slopes between the curves for
the samples HG1 and HG2 and between the HG2 curves are caused
primarily by small, different misalignments between the external
applied magnetic field and the sample main plane. After subtraction
of the dominating linear background small ferromagnetic loops remain,
see Fig.~2(b). Ferromagnetic-like loops are also seen in the virgin
samples \cite{pabloprb02}. The proton irradiation performed at low
temperatures causes a clear enhancement of the ferromagnetic effect
in relation to that of the virgin sample. In contrast, the
Fe-implantation does not lead to a detectable increase of the
ferromagnetism. Even a Fe-implantation with a tenfold dose (sample
HG2$_-$Fe5.0) gives no enhancement of the ferromagnetic contribution.

The ferromagnetic behaviour of the irradiated samples is better
demonstrated in Fig.~3 where the magnetization curves of the samples
HG1 and HG2 are plotted after subtraction of the contributions of the
corresponding virgin samples. To show the influence of irradiation
the magnetization in Fig.~3 is related to the mass of the irradiated
sample region. It is known that the ferromagnetically active layer in
a proton-irradiated graphite sample is very small in comparison to
the penetration depth of the protons for the fluence and energy used
here \cite{barzola2}. For the determination of the irradiated area of
sample HG1 we assumed a spot radius of $2.6~\mu$m and a thickness of
the ferromagnetic part of 1~$\mu$m. After subtraction of the
paramagnetic part we get a value of 3~emu/g at $T = 300$~K for the
ferromagnetic saturation magnetization caused by the proton
irradiation. If we assume a larger area due to irradiation
broadening, i.e. 2.5~mm$^2$, we estimate a magnetization at
saturation of the order of 0.6~emu/g.

Concluding, for similar amount of irradiated ions under similar
energies our results show clearly that only the proton-irradiation
induces a ferromagnetic contribution under the conditions used in
this study. In comparison with proton-irradiation in the
Fe-implanted samples the structural disorder, i.e. vacancy
density, produced by the bombardment appears to be too large to
induce ferromagnetic correlations. A large disorder in the atomic
lattice may prevent the existence of a long range magnetic
interaction through the graphene structure \cite{yaz07,pis07}. We
speculate that the ferromagnetism in the proton-irradiated sample
is a combination of the disorder plus the action of hydrogen ions
(already in the sample very probably as $H_2$ and dissociated by
the irradiation). Independently done electric force measurements
and scanning transmission x-ray microscopy on magnetic spots
provide further evidence that the irradiation induces  a change in
the electronic $\pi$-band and $\sigma-$band, leaving the
ferromagnetic phase with a smaller conduction electron density at
Fermi level \cite{sch07}, a result compatible with recently done
theoretical work \cite{pis07} that indicates that the
ferromagnetic graphitic phase should have insulating properties.

 We gratefully acknowledge discussions with N. Garc\'ia, M. A.
Vozmediano, L. Pisani and N. Harrison and T. Butz for the
permanent support. This work was done in the framework of the EU
project "Ferrocarbon" and partially supported by the DFG under ES
86/11.

%\bibliographystyle{apsrev}
%\bibliography{D:/data/hopg/magnetic_carbon}

\end{document}